\documentclass[12pt]{article}

\usepackage[dvips]{epsfig}
\usepackage{latexsym}
\usepackage{graphicx}
\usepackage[bf,footnotesize]{caption}   
\begin{document}
\setlength{\unitlength}{1mm}

\newcommand{\be}{\begin{equation}}
\newcommand{\ee}{\end{equation}}
\newcommand{\ba}{\begin{eqnarray}}
\newcommand{\ea}{\end{eqnarray}}
\newcommand{\ban}{\begin{eqnarray*}}
\newcommand{\ean}{\end{eqnarray*}}

\newcommand{\e}{{\mathrm e}}
\newcommand{\bbox}{\bar{\phantom{!}\Box\phantom{!}}}
\newcommand{\Tr}{\phantom{!}\mbox{\bf Tr}\phantom{!}}
\newcommand{\tr}{\phantom{!}\mbox{\bf tr}\phantom{!}}
\newcommand{\ra}{\rangle}
\newcommand{\la}{\langle}

\newcommand{\n}[1]{\label{#1}}
\newcommand{\eq}[1]{Eq.(\ref{#1})}
\newcommand{\ind}[1]{\mbox{\bf\tiny{#1}}}
\renewcommand\theequation{\thesection.\arabic{equation}}

\newcommand{\nn}{\nonumber \\ \nonumber \\}
\newcommand{\nl}{\\  \nonumber \\}
\newcommand{\pr}{\partial}
\renewcommand{\vec}[1]{\mbox{\boldmath$#1$}}

\title{{\hfill {\small Alberta-Thy-01-03} } \vspace*{2cm} \\
Particle and light motion in a space-time of a five-dimensional
rotating black hole}
\author{
Valeri Frolov\footnote{e-mail: frolov@phys.ualberta.ca}
 and
Dejan Stojkovi\'{c}\footnote{e-mail: dstojkov@phys.ualberta.ca}
}
\maketitle
\noindent
\centerline{ \em
Theoretical Physics Institute, Department of Physics,}
\centerline{ \em University of Alberta,  Edmonton, Canada T6G 2J1}
\bigskip

\noindent

\begin{abstract} We study motion of particles and light in a
space-time of a 5-dimensional rotating black hole. We demonstrate
that the Myers-Perry metric describing such a black hole in addition
to three Killing vectors possesses also a Killing tensor. As a result,
the Hamilton-Jacobi equations of motion allow a separation of
variables. Using first integrals  we present the equations of motion
in the first-order form. We describe different types of motion of
particles and light and study some interesting special cases.
We proved that there are no stable circular orbits in equatorial planes
in the background of this metric.
\end{abstract}

\vspace{.3cm}


\section{Introduction}
\setcounter{equation}0

Brane world models with large extra dimensions have recently attracted a lot of
interest \cite{BW}. An important generic feature of these models is that the
fundamental quantum gravity scale may be very low  (of order TeV) and the
size of extra spatial dimensions may be much larger than the Planck
length ($\sim 10^{-33}$cm). In the models with `large extra-dimensions'
the minimal mass of black hole can be also much smaller that the Planck
mass ($10^{16}$ GeV) and be of order TeV. Such mini-black
holes could be produced in particle collisions in near future
colliders  and in cosmic ray experiments \cite{ACC_CR}. One can expect
that majority of black holes produced in such way would be rotating
\cite{ACC_CR,FrSt}. Estimations show that such higher dimensional
mini-black holes can be described within the classical solutions of
vacuum Einstein's equations. These mini black holes can be attached to
the brane or, if the brane tension is small, can leave the brane and
travel in the bulk space \cite{FrSt}.  Recently, the problem of higher
dimensional black holes has  attracted a lot of attention \cite{citations}.

Higher dimensional black holes were also studied in string
theory. Motivated by ideas arising in various models in string
theory a lot of work was done studying supersymmetric higher dimensional black holes
especially in 5-dimensional space-time. For example,
supersymmetric rotating black holes were studied in \cite{GibHer}
and  \cite{Her}. The solution analyzed there is not a vacuum solution
of Einstein's equations and requires some special choice of
parameters in order accommodate the supersymmetry. In \cite{Cvetic},
rotating black holes were studied in the context of string theory.
Although the solution described there is very general (a boosted
vacuum solution of  Einstein's equations) the authors mainly
concentrate on scalar field gray body factors. In the previous paper
\cite{K5D}, we studied some general geometrical properties of a
five-dimensional rotating black hole and the propagation of
five-dimensional massless scalar field in the background of such a
black
hole.

In this paper, we extend our analysis to properties of motion of
particles and light in the space-time of a five-dimensional rotating black
hole. We demonstrate the existence of a Killing tensor in such a
space-time. We describe different types of motion of particles and
light and study some interesting special cases. The five-dimensional
metric is algebraically special and allows two families of principle null
congruences. These congruences are geodesic but not shear free.
We also showed that there are no stable circular orbits in equatorial planes
in the background of this metric.

\section{Myers-Perry metric and its properties}\label{s1}
\setcounter{equation}0

The metric for five-dimensional rotating black hole\footnote{Note that the
Myers-Perry solution is not the only black hole solution of the five dimensional
vacuum Einstein equations. For example see \cite{other}.} in the
Boyer-Lindquist coordinates is (\cite{MyPe:86}):

\be \label{Kerr5D} ds^2={\rho^2\over 4\Delta}\, dx^2+\rho^2\,
d\theta^2\, - dt^2 +(x+a^2)\, \sin^2\theta\, d\phi^2+ (x+b^2)\,
\cos^2\theta\, d\psi^2   \ee
\be + {r_0^2\over \rho^2} \left[dt+a\, \sin^2\theta\, d\phi +b\,
\cos^2\theta\, d\psi  \right]^2\, . \ee
Here,
\be
\rho^2=x+a^2\,\cos^2\theta+b^2\,\sin^2\theta\, ,
\ee
\be
\Delta=(x+a^2)(x+b^2)-r_0^2\, x\, .
\ee
Angles $\phi$ and $\psi$ take values from the interval
$\left[0,2\pi \right]$, while angle $\theta$ takes values from
$\left[0,\pi/2 \right]$. Note also that
instead of the `radius' $r$ we use the coordinate $x=r^2$.
This will allow us to simplify calculations and make many of the
expressions more compact.

The black hole horizon is located at $x=x_+$ where
\be
x_{\pm}={1\over 2}\left[r_0^2-a^2-b^2\pm
\sqrt{(r_0^2-a^2-b^2)^2-4a^2b^2}\right]\, .
\ee
The angular velocities $\Omega_a$ and $\Omega_b$ and the surface
gravity $\kappa$ are
\be \Omega_a={a\over x_+ +a^2}\, , \hspace{1cm} \Omega_b={b\over
x_+ +b^2}\, ,\hspace{1cm} \kappa= \left. {\partial_x\Pi -r_0^2\, \over r_0^2\,
\sqrt{x} }\right|_{x=x_+} \, . \ee

For the metric (\ref{Kerr5D})
\be \sqrt{-g}=\frac{1}{2} \sin\theta\cos\theta\, \rho^2\, . \ee
We shall also need the following expressions for the contravariant
components of the metric
\[
g^{tt}=\frac{1}{\rho^2 } \left[(a^2-b^2)\,\sin^2\theta -
{(x+a^2)[\Delta+r_0^2(x+b^2)]\over \Delta} \right]\, ,
\]
\[
g^{t\phi}=\frac{ar_0^2(x+b^2)}{\rho^2 \Delta}\, ,\hspace{1cm}
g^{t\psi}=\frac{br_0^2(x+a^2)}{\rho^2 \Delta}\, ,
\]
\be
g^{\phi\phi}=\frac{1}{\rho^2} \left[{1\over
\sin^2\theta}-{(a^2-b^2)(x+b^2)+b^2r_0^2\over \Delta}\right] \, ,
\ee
\[
g^{\psi\psi}=\frac{1}{\rho^2} \left[{1\over
\cos^2\theta}+{(a^2-b^2)(x+a^2)-a^2r_0^2\over \Delta} \right]\, ,
\]
\[
g^{\phi \psi}=-\frac{abr_0^2}{\rho^2 \Delta}\, ,\hspace{1cm}
g^{xx}=4\frac{\Delta}{\rho^2}\, ,\hspace{1cm}
g^{\theta \theta}=\frac{1}{\rho^2}\, .
\]

The metric (\ref{Kerr5D}) is invariant under the following
transformation
\be \n{tran}
a \leftrightarrow b\, ,\hspace{0.5cm} \theta \leftrightarrow
\left( {\pi\over 2}-\theta \right)\, ,\hspace{0.5cm} \phi
\leftrightarrow \psi \, . \ee
It possesses 3 Killing vectors, $\partial_t$, $\partial_{\phi}$
and $\partial_{\psi}$. For $a=b$ the metric has 2 additional Killing
vectors  \cite{Galtsov,K5D}:
\be \cos\, \partial_{\bar{\theta}}-\cot\bar{\theta}\, \sin\bar{\phi}\,
\partial_{\bar{\phi}}+{\sin \bar{\phi}\over \sin\bar{\theta}}\, \partial_{\bar{\psi}}\,
\ee
and
\be
-\sin\bar{\phi}\, \partial_{\bar{\theta}}-\cot\bar{\theta}\,
\cos\bar{\phi}\,
\partial_{\bar{\phi}}+{\cos\bar{\phi}\over \sin\bar{\theta}}\, \partial_{\bar{\psi}}\, .
\ee
where $\bar{\phi}=\psi-\phi$, $\bar{\psi} =\psi+\phi $ and $\bar{\theta} =2 \theta$

In the next section we shall demonstrate that
in the general case the metric (\ref{Kerr5D}) has also the Killing
tensor $K^{\mu\nu}$ satisfying the equation
\be\label{KE}
K_{(\mu\nu ; \sigma)}=0
\ee

\section{Equations of motion for particles and light. First integrals.}
\label{s2}
\setcounter{equation}0

The equations of motion of a test particle of mass $m$ in a
curved space-time given by a metric $g_{\mu \nu}$ are:
\be \n{3.1}
\frac{D^2 x^\mu}{D\tau^2} =0 \, ,
\ee
where $\frac{D }{D\tau}$ denotes covariant derivative with respect to
proper time $\tau$.  In this section we study these equations in the
Myers-Perry metric (\ref{Kerr5D}) by using the Hamilton-Jacobi
method. This consideration is similar to the approach developed by
Carter for 4dimensional Kerr metric \cite{Carter} (see also
\cite{MTW}).

The equations (\ref{3.1}) can be derived from the Lagrangian:
\be \label{L} L = \frac{1}{2} g_{\mu \nu} \dot{x^\mu} \dot{x^\nu}\, ,
\ee
where a dot denotes partial derivative with respect to an affine
parameter $\lambda$. For consistency, we chose
\be \label{tau}  \tau = m \lambda \, ,\ee
which is equivalent to
\be  g_{\mu \nu} \dot{x^\mu} \dot{x^\nu} = -m^2 \, .\ee
The conjugate momenta following from (\ref{L}) are
\be p_\mu=g_{\mu \nu} \dot{x^\nu} \, . \ee

Thus, the Hamiltonian is
\be \label{H} H = \frac{1}{2} g^{\mu \nu} p_\mu p_\nu \, .\ee
The momenta calculated for the metric (\ref{Kerr5D}) are:
\begin{eqnarray}
&& p_t = \left( -1+ \frac{r_0^2}{\rho^2}  \right) \dot{t}
+\frac{r_0^2}{\rho^2} \dot{\phi} +\frac{r_0^2 b \cos^2 \theta
}{\rho^2}  \dot{\psi} \, , \\
&& p_\phi = \frac{r_0^2a\sin^2\theta}{\rho^2}  \dot{t} +\left(
x+a^2+
\frac{r_0^2 a^2 \sin^2 \theta}{\rho^2} \right) \sin^2 \theta
\dot{\phi} +\frac{r_0^2 a b \sin^2 \theta \cos^2 \theta
}{\rho^2}  \dot{\psi} \, ,\nonumber \\
&& p_\psi = \frac{r_0^2 b \cos^2 \theta}{\rho^2}  \dot{t}
+\frac{r_0^2 a b \sin^2 \theta \cos^2 \theta }{\rho^2}
\dot{\phi}+\left( x+b^2+ \frac{r_0^2 b^2 \cos^2 \theta}{\rho^2}
\right) \cos^2 \theta \dot{\psi} \, , \nonumber \\
&& p_x = \frac{\rho^2}{4\Delta} \dot{x} \, , \nonumber \\
&& p_\theta = \rho^2 \dot{\theta} \, .\nonumber
\end{eqnarray}

From the symmetries of the metric (\ref{Kerr5D}) it follows that
at least three constants of motion should exist. They correspond
to conservation of energy, $E$, and angular momenta in two
independent planes defined by angles $\phi$ and $\psi$. Thus,
\be p_t = -E \, , \ \ \ \ \  p_\phi = \Phi \ \ \ \ \ {\rm and} \
\ \ \ \  p_\psi = \Psi \, . \ee

We also have the constant of motion corresponding to conservation
of rest mass which is given by (\ref{tau}).

In order to solve the system of equations of motion completely,
we need one more constant of motion. This can be obtained using
Hamilton-Jacobi method of solving equations of motion.

From the Hamiltonian (\ref{H}) we have Hamilton-Jacobi equations
in  the form:
\be \label{HJeq} -\frac{\partial S}{\partial
\lambda} = H =
\frac{1}{2} g^{\mu \nu} \frac{\partial S}{\partial x^\mu}
\frac{\partial S}{\partial x^\nu} \ee
where $S$ is the Hamilton-Jacobi action. Since it was proven in
\cite{K5D} that the equation of motion can be separated, the
action must take the form:
\be \label{HJac} S= \frac{1}{2} m^2 \lambda -Et+ \Phi\phi +
\Psi\psi + S_\theta + S_x \ee where $S_\theta$ and $S_x$ are
functions of $\theta$ and $x$ respectively. From equations
(\ref{HJeq}) and (\ref{HJac}) we can conclude:
\be     \label{thetaeq}
 \left( \frac{\partial S_\theta}{\partial \theta} \right)^2 +
(m^2-E^2) (a^2 \cos^2\theta+ b^2 \sin^2\theta ) +{1\over
\sin^2\theta}\Phi^2+ {1\over \cos^2\theta}\Psi^2 = K\, ,
\ee
and
\be   \label{xeq}
 4 \Delta \left(
\frac{\partial S_x}{\partial x} \right)^2 + (m^2-E^2)x -
\frac{r_0^2(x+a^2)(x+b^2)}{\Delta} \, {\cal E}^2
 -(a^2-b^2)\left({ \Phi^2 \over (x+a^2)}
  -{\Psi^2\over (x+b^2)}\right)    =-K \, ,
\ee
where
\be\n{EE0}
{\cal E}= E+ \frac{a \Phi}{x+a^2}+\frac{b \Psi}{x+b^2}\, .
\ee
Here $K$ is a new constant of the motion. We used the freedom
$K \rightarrow K + {\rm const}$ (in this case const $=a^2E^2$)
to obtain  the equations in the form invariant under the
transformation (\ref{tran}). Using relations  $ p_\theta
=S_{,\theta}$ and   $ p_x =S_{,x}$ we can replace
$(S_\theta)_{,\theta}$ and $(S_x)_{,x}$  by $p_\theta$ and $p_x$
respectively.

The conserved quantity $K$ is of the second order in the momenta
$p_{\mu}$ and it is
related to the Killing tensor $K_{\mu\nu}$ as follows
\be\label{K}
 K^{\mu\nu} p_\mu p_\nu =K \, ,
\ee
By comparing (\ref{thetaeq}) with (\ref{K}) one obtains
\be  \label{kt}
K^{\mu\nu} =-(a^2\cos^2\theta +b^2\sin^2\theta )(g^{\mu\nu}+
\delta^\mu_t \delta^\nu_t) +
   \frac{1}{\sin^2\theta}\delta^\mu_\phi
   \delta^\nu_\phi+\frac{1}{\cos^2\theta}
   \delta^\mu_\psi \delta^\nu_\psi+\delta^\mu_\theta
   \delta^\nu_\theta\, .
\ee
A similar result for the 4-dimensional Kerr metric was obtained by
Carter in 1968 \cite{Carter}. We used GRTensor program to check
directly that $K^{\mu\nu}$ does obey the equation (\ref{KE}).

Equations (\ref{thetaeq}) and (\ref{xeq}) can be written in a
compact form:
\be\n{eqs}
\frac{\partial S_\theta}{\partial \theta}  = \sigma_{\theta} \sqrt{\Theta}\, ,
\hspace{1cm}
\frac{\partial S_x}{\partial x} = \sigma_{x} \sqrt{X} \, .
\ee
Here functions $\Theta$ and $X$ are given by
\be\n{Theta}
\Theta =(E^2-m^2) (a^2 \cos^2\theta+ b^2 \sin^2\theta )
-{1\over \sin^2\theta}\Phi^2- {1\over \cos^2\theta}\Psi^2 + K \, ,
\ee
\be\n{XX}
X= {{\cal X}\over4 \Delta^2} \, ,
\ee
\be\n{X}
{\cal X}= \Delta \left[ x (E^2-m^2)
 +(a^2-b^2)\left({\Phi^2 \over  (x+a^2)}
  -{\Psi^2\over  (x+b^2)}\right)  -  K \right]
+r_0^2(x+a^2)(x+b^2)\, {\cal E}^2 \, .
\ee
The sign functions $\sigma_{\theta}=\pm$ and $\sigma_{x}=\pm$ in the
right hand sides of two equations (\ref{eqs}) are independent from
one another. In each of the equations the change of the sign occurs
at a turning point where the expression in the right hand side
vanishes.

We can write the Hamilton-Jacobi action in terms of these
functions:
\be
S=\frac{1}{2} m^2 \lambda -Et+ \Phi\phi + \Psi\psi +\sigma_{\theta} \int^\theta
\sqrt{\Theta} d\theta +\sigma_{x}\int^x \sqrt{X}dx \, .
\ee
By differentiating with respect to $K$, $m$, $E$, $\Phi$ and
$\Psi$ we obtain the  solution of the Hamiltonian-Jacobi
equations as
\begin{eqnarray}    \label{solutions}
&& \int^{\theta} \frac{d\theta}{ \sqrt{\Theta}} =
\int^{x} \frac{dx}{ 4\Delta \sqrt{X}} \, ,   \\
&& \lambda = \int^{\theta} \frac{1}{ \sqrt{\Theta}} ( a^2\cos^2
\theta + b^2 \sin^2 \theta ) d\theta+
\int^{x}  \frac{1}{ \sqrt{X}}  \frac{x}{4\Delta}  dx \, ,  \\
&& t = \int^{\theta} \frac{1}{ \sqrt{\Theta}}  E (a^2\cos^2
\theta + b^2 \sin^2 \theta ) d\theta +  \nonumber \\
&& \int^{x} \frac{1}{ \sqrt{X}}  \left[
\frac{r_0^2(x+a^2)(x+b^2)}{4\Delta^2} {\cal E} +
\frac{xE}{4\Delta}\right] dx \, , \\
&& \phi = \int^{\theta}  \frac{1}{ \sqrt{\Theta}}
\frac{\Phi}{\sin^2 \theta}  d\theta - \int^{x}  \frac{1}{
\sqrt{X}} \left[ \frac{a r_0^2(x+b^2)}{4\Delta^2}\, {\cal E}
 +\frac{(a^2-b^2)\Phi}{4\Delta (x+a^2)}   \right] dx \, ,
  \\
&& \psi = \int^{\theta}  \frac{1}{ \sqrt{\Theta}}
\frac{\Psi}{\cos^2 \theta}  d\theta - \int^{x} \frac{1}{
\sqrt{X}} \left[ \frac{br_0^2(x+a^2)}{4\Delta^2}\, {\cal E}
-\frac{(a^2-b^2)\Psi}{4\Delta (x+b^2)}   \right] dx \, .
\end{eqnarray}

Often  it is more convenient to rewrite these equations in
the form of the first order differential equations
\be
\rho^2 \dot{\theta} = \sigma_{\theta} \sqrt{\Theta}\, ,
\ee
\be\n{xdot}
\rho^2 \dot{x} = \sigma_x 2 \sqrt{{\cal X}}\, ,
\ee
\be
\rho^2 \dot{t} =   E \rho^2
 + \frac{r_0^2(x+a^2)(x+b^2)}{\Delta}\, {\cal E}\, ,
\ee

\be
\rho^2 \dot{\phi} =  \frac{\Phi}{\sin^2 \theta}   -  \frac{a
r_0^2(x+b^2)}{\Delta}\, {\cal E}-\frac{(a^2-b^2)\Phi}{(x+a^2)}\, ,
\ee

\be
\rho^2 \dot{\psi} = \frac{\Psi}{\cos^2 \theta}  -
\frac{br_0^2(x+a^2)}{\Delta}\, {\cal E}+\frac{(a^2-b^2)\Psi}{(x+b^2)}  \, .
\ee
These equations can be obtained from  equations (\ref{solutions})
by direct differentiation with respect to the affine parameter
$\lambda$.

\section{General types of motion and special cases}
\label{s3}
\setcounter{equation}0

\subsection{Radial motion}

The geodesic world line of a particle in the Myers-Perry metric is
completely determined by the first integrals of motion $E$, $\Phi$,
$\Psi$, and $K$. We discuss the motion in the black hole exterior, so
we assume that $\Delta>0$. Consider ${\cal X}$ given by (\ref{X}) as
the function of $x$ for fixed values of the other parameters. At
large distances the leading term of $X$ contains the factor
$E^2-m^2$. For $E^2<m^2$ the function $X$ becomes negative at large
$x$. Hence only orbits with $E^2>m^2$ can extend to infinity. These
orbits are called {\em unbound}. For $E^2<m^2$ the orbit is always
{\em bounded}, that is the particle cannot reach infinity.

To study qualitative characteristics of the motion of test particles
in the metric (\ref{Kerr5D}), it is convenient to use the {\em effective
potential}. Let us write ${\cal X}$ as
\be
{\cal X}=\alpha\, E^2-2\beta\, E+\gamma\, ,
\ee
where
\be
\alpha=  \Delta x+r_0^2(x+a^2)(x+b^2)\, ,
\ee
\be
\beta = -r_0^2(x+a^2)(x+b^2)\left(
\frac{a\Phi}{x+a^2}+\frac{b\Psi}{x+b^2} \right)\, ,
\ee
\be
\gamma = \Delta \left[-m^2x+(a^2-b^2)\left(\frac{\Phi^2}{x+a^2}-
\frac{\Psi^2}{x+b^2}\right)-K \right]
+r_0^2(x+a^2)(x+b^2)\left[\frac{a\Phi}{x+a^2}+\frac{b\Psi}{x+b^2}\, .
\right]^2
\ee

The radial turning points ${\cal X}=0$ are defined by the condition
$E=V_{\pm}(x)$, where
\be\n{ef_pot}
V_{\pm}={\beta\pm \sqrt{\beta^2-\alpha\gamma}\over \alpha}\, .
\ee
The quantities $V_{\pm}$ are called the {\em effective potentials}. They
are functions of $x$, and integrals of motion  $\Phi$,
$\Psi$, and $K$.

The limiting values of the effective potentials $V_{\pm}$ at infinity
and at the horizon are
\be
V_{\pm}(x=\infty)= \pm m\, ,\hspace{1cm}
V_{\pm}(x_+)=\Phi\, \Omega_a+\Psi\, \Omega_b\, .
\ee
The motion of a particle with energy $E$ is possible only in the
regions where either $E\ge V_+$ or $E\le V_-$. Expression
(\ref{ef_pot}) remains invariant under transformations $E\to -E$,
$\Phi\to -\Phi$, $\Psi\to -\Psi$ relating these regions. In the
absence of rotation, $a=b=0$ the second region, $E\le V_-$ is
excluded.

\subsection{Motion in $\theta$-direction}

Consider now motion in the $\theta$-direction. Since $\Theta\ge 0$,
bounded motion with $E^2<m^2$ is possible only if $K\ge 0$. For
bounded motion and $a\ne 0$, $b\ne 0$, \ \ $K>0$. For $K>0$ there
exist both bounded and unbounded trajectories. A particle can reach
subspace $\theta=0$ only if $\Phi=0$, and the subspace $\theta=\pi/2$
only if $\Psi=0$. For $\Psi=0$ the orbit is in the $\theta=\pi/2$
plane if $K=\Phi^2-(E^2-m^2)b^2$. Similarly, for $\Phi=0$ the orbit
is in the $\theta=0$ plane if $K=\Psi^2-(E^2-m^2)a^2$.

Special class of motion is a case when particles are moving
quasi-radially along trajectories on which the value of the angle
$\theta$ remains constant, $\theta=\theta_0$. The relation between the
integrals of motion corresponding to this type of motion can be found
by solving simultaneously the equations
\be
\Theta(\theta_0)={d\Theta\over d\theta}|_{\theta_0}=0\, .
\ee
These equations are of the form
\be\n{THETA}
(E^2-m^2) (a^2 \cos^2\theta_0+ b^2 \sin^2\theta_0 )
-{\Phi^2\over \sin^2\theta_0}- {\Psi^2\over \cos^2\theta_0} + K =0\,
,
\ee
\be\n{THETA'}
(E^2-m^2) (b^2-a^2)+{\Phi^2\over \sin^4\theta_0}- {\Psi^2\over
\cos^4\theta_0}=0\, .
\ee
In the second equation we excluded cases $\theta_0=0$ and
$\theta_0=\pi/2$.

\subsection{Circular orbits}

Characteristic property of the 4-dimensional gravitational field is
the existence of bounded orbits located in the exterior of the black
hole. For such orbits there are 2 turning points, corresponding to the
minimum and maximum of the radial coordinate for the particle
trajectory. It means that the line $E$=const crosses the effective
potential $V(r)$ curve at 2 points, $r_1$ and $r_2$. Between these
points $E>V(r)$. The function $V(r)$ has minimum at $r_1<r_{min}<r_2$.
For the special case when $E=V(r_{min})$ the orbit is circular. Thus
the existence of stable circular orbits is a necessary condition for
the existence of bounded orbits located in the exterior of the black
hole.

It is well known that  there are no stable circular orbits
in the Newtonian gravity in a space-time with more that three spatial
dimensions. This is also true in the general relativity. Namely the
Schwarzschild metric
\be
ds^2=-F\, dt^2+{dr^2\over F}+r^2\, d\Omega_{k+2}^2\, ,\hspace{0.5cm}
F=1-\left({r_0\over r}\right)^{k+1}\,
\ee
where $d\Omega_{k+2}^2$ is the metric on a unit $(k+2)$-dimensional
sphere, does not allow stable circular orbits. The effective potential
for the radial motion of a particle with mass $m$ in this field is
\be
V_S^2(r) = -F \left( 1+\frac{L^2}{m^2 r^2} \right)
\ee
where $L$ is the total angular momentum. It is easy to see that this
potential does not have any minima in the interval
$(r_0,\infty)$.

With a change of variables $y = r_0^2/r^2$ ($y$ goes from 0 to 1) we can write:

\be
V_S^2(y) = (1-y^\frac{1+k}{2})(1+\frac{L^2}{r_0^2 m^2}y)
\ee

The second derivative of the function $V_S^2(y)$ with respect to $y$ is:
\be \label{secder}
\frac{\partial^2 \left( V_S^2(y)\right)}{\partial y^2}=-\frac{k^2-1}{4} y^\frac{k-3}{2}-\frac{(k+3)(k+1)L^2}{4r_0^2m^2}
y^\frac{k-1}{2}
\ee
This function is zero only for $y \leq 0$ which means that $V_S^2$
can not have more than one extremum for positive $y$. Note also that
$V_S^2(y)$ takes value 1 at $y=0$ and value 0 at $y=1$. Since
$V_S^2(y)$ is
non-negative, we conclude that it is either monotonically decreasing
function or it the has one maximum at this interval without possibility for
having a minimum. The new variable $y$ is a monotonic function of
$r$ and  the original function $V_S^2(r)$ (and therefore
$V_S(r)$) does not have a minimum either. Thus, the potential
$V_S(r)$ does not allow any motion which is confined in a finite
interval of the radial coordinate.
Similar conclusion can be derived if we consider an effective potential
for angular motion \cite{Tan}.

\begin{figure}[h!]
\center
\epsfig{file=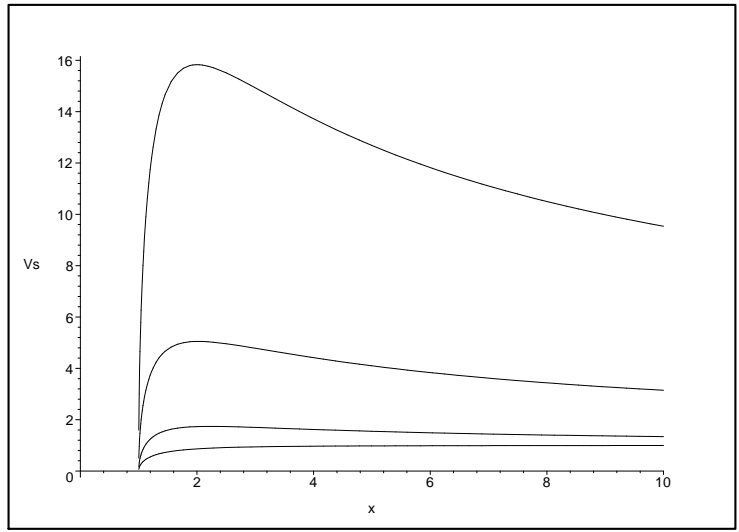, width=7cm, angle=-90}
\caption{The  effective potential $V_S$ for radial motion in the five-dimensional
Schwarzschild space-time for several values of
the angular momentum $L$.
We set $m=r_0=1$ and four different curves correspond to $L^2=1, 10,
100 {\ \rm and \ } 1000$  from the bottom up respectively.}
\label{VpS}
\end{figure}

Figure~\ref{VpS} shows a plot of the potential $V_S$ for the 5-dimensional
Schwarzschild space-time for several values of
the angular momentum $L$.

We shall prove now that the stable circular orbits do not exist in
the Myers-Perry metric at least in the case when this orbits are in
the `equatorial' planes, $\theta=0$ and $\theta=\pi/2$. We give
the proof for the case $\theta=\pi/2$ and the other case
is similar.

Circular orbits $x=x_0=$const are defined by the equations
\be \label{sco}
{\cal X} = 0\, ,\hspace{0.5cm}
\frac{\partial {\cal X}}{\partial x} = 0\, ,
\ee
where ${\cal X}$ is given by (\ref{XX}). For orbits in the plane
$\theta=\pi/2$ one has $\Psi=0$ and  $K=\Phi^2-(E^2-m^2)b^2$ (see
previous subsection). Substituting these relations into eq.
(\ref{sco}) we obtain a system of quadratic
equations\footnote{This method of analyzing circular orbits is
similar to the method used by Bardeen, Press, and Teukolsky
\cite{BPT} for four-dimensional Kerr metric.} for
variables $E$ and $\Phi$. The solutions are
\be
E_{\sigma}=\sigma E\, ,\hspace{0.5cm}\Phi_{\sigma}=\sigma \Phi\, ,
\hspace{0.5cm}\sigma=\pm\, ,
\ee
\be\n{EE}
{E\over m} = \frac{y}
{\sqrt{y^2-y_0^2}}\, ,
\ee
\be\n{FF}
{\Phi\over m} =\frac{(y+a^2-b^2-r_0^2)r_0}
{\sqrt{y^2-y_0^2}}\, .
\ee
where
\be
y=x\mp r_0 a +b^2-r_0^2\, ,\hspace{0.5cm}
y_0=r_0\sqrt{ (r_0\pm a)^2-b^2}\, .
\ee

Since the constants $E$ and $\Phi$ are to be real, the expression
under the root square in the denominators of (\ref{EE}) and (\ref{FF}) are
to be positive. This happens when
\be\n{sq1}
y^2>y_0^2\,.
\ee
But for these values of $y$ we have $E^2>m^2$ and the potential must
allow unbounded motion.
This means that if the minimum of the effective potential exists, there must be also
at least one maximum. But, the equation for the second derivative
\be
\frac{\partial^2 {\cal X}}{\partial x^2} =2\left[ (E^2-m^2)(3x+a^2+2b^2)
-\Phi^2+r0^2m^2 \right]= 0\, ,
\ee
has only one solution for $x$ and this situation is impossible.
Thus, we  conclude that there
are no stable circular orbits around the rotating five dimensional
Myers-Perry black hole, at least in the `equatorial' planes.

One may conjecture that the absence or bounded stable orbits in the black
hole exterior is a generic property of higher dimensional black holes.

\subsection{Principal null congruences}

We consider now null rays moving along $\theta_0=$const surfaces. One
can use conditions (\ref{THETA}) and (\ref{THETA'}) (with $m=0$) to
determine two of the constants of motion. After this a general
solution is specified by $E$, $K$, $\theta_0$, and $t_0$, $\phi_0$,
$\psi_0$. The last 3 constants are required as initial data for 3
Killing variables, $t$, $\phi$, and $\psi$. Let us note that the
parameter $E$ is not important. It can be `gauged away' be rescaling
of the affine parameter $\lambda \to E\lambda$. Thus, we have
five-parameter family of null geodesics. In order to have a
congruence of null geodesics one needs to fix one more parameter. The
following special choice is very convenient

\be
\Phi=-E a \sin^2\theta_0\, ,\hspace{1cm}
\Psi=-E b \cos^2\theta_0\, .
\ee
For this choice the equation (\ref{THETA'}) is automatically
satisfied, while the equation (\ref{THETA}) gives
\be
K=E^2\, (a^2-b^2)\, (\sin^2\theta_0-\cos^2\theta_0)\, .
\ee
The null geodesics from this congruence are uniquely specified by the
parameters $\theta_0$, $t_0$, $\phi_0$ and $\psi_0$.
The null vectors tangent to the geodesics from this congruences
are
\be\n{PNG}
L_{\pm}^{\mu}\partial_{\mu}={(x+a^2)(x+b^2)\over \Delta}\left[ \partial_t
-{a\over x+a^2}\partial_{\phi} -{b\over x+b^2}\partial_{\psi}\right] \pm
2\sqrt{x}\partial_x\, .
\ee
The two different congruences differ by the choice of sign in
(\ref{xdot}). By  analogy with similar congruences in the
four-dimensional Kerr
geometry, we call the congruences generated by $L_{\pm}^{\mu}$
{\em principal null congruences}.

By using GRTensor program one can check that both of the null principal
congruences obey a condition
\be
{L_{\pm}}_{[\alpha}\, C_{\beta]\gamma\delta \epsilon}\,
{L_{\pm}}^{\gamma}\,L_{\pm}^{\delta}\, =0\, ,
\ee
where $C_{\beta\gamma\delta \epsilon }$ is the Weyl tensor. In 4
dimensional case a similar condition means that the space-time is
algebraically special and belong to the Petrov class $D$ (or more
degenerate)\footnote{Petrov classification for five-dimensional
metrics of Euclidian signature was given in \cite{Smet}. Classification of higher
dimensional space-times and its relation to the existence of principal
null geodesics
is an interesting problem. In our paper we use the term
``algebraically special " only in the sense that the space-time
allows  principal null geodesics.}.

One can check that the shear $\sigma_{\pm}$ defined by (\ref{a.6}) for
$L_{\pm}^{\gamma}$ is
\be\n{sigma}
\sigma_{\pm}=\sqrt{2\over 3}{(a^2\cos^2\theta +b^2\sin^2\theta)\over
 \rho^2 \sqrt{x} }\, ,
\ee
and it does not vanish\footnote{ In the 4-dimensional case the
following theorem proved by Goldberg and Sachs \cite{GS} is valid: If
a vacuum gravitational field is algebraically special then principal
null congruences must be shear free. The result
(\ref{sigma}) indicates that this theorem might not be valid in
the 5-dimensional case.}. To perform the calculation of $\sigma$ we used GRTensor program.
The direct calculations based on relation (\ref{a.5}) of the Appendix
result in very long expressions, so that the standard GRTensor
simplification procedure does not work. We found that calculations
based on the second form of the expression for $\sigma$ in
(\ref{a.6}) are much shorter and allows one to arrive to the result
(\ref{sigma}) much faster.

The principal null geodesic congruences (\ref{PNG}) were used in
\cite{MyPe:86} to establish relations between Boyer-Lindquist and
Kerr-Schild coordinates for the Myers-Perry metric.

Similarly to the 4-dimensional Kerr metric both of the null
principle vectors $L_{\pm}^{\mu}$ are eigen-vectors of the  tensor
$\xi_{(t)\mu;\nu}$, where
$\xi_{(t)}^{\mu}\partial_{\mu}=\partial_t$
is the Killing vector which is time-like at infinity. Namely the
following relations are valid
\be
\xi_{(t)\mu;\nu }\, L_{\pm}^{\nu}=\kappa_{\pm}\, {L_{\pm}}_{\mu}\, ,
\ee
where
\be
\kappa_{\pm}= \pm\, {\sqrt{x}\, r_0^2\over \rho^4}\, .
\ee

\section{Conclusions}
\label{C}
\setcounter{equation}0

We discussed motion of particles and light in the space-time of
five-dimensional rotating black hole. There is an intriguing
similarity of this problem with the case of the four-dimensional Kerr
metric. In both cases the Hamilton-Jacobi equations for particles and
light allow separation of variables. This property follows from the
existence of a Killing tensor in addition to the Killing vectors.

We described different types of motion of particles and light in the
background of five-dimensional rotating black hole including some
interesting special cases (like radial motion and motion with constant $\theta$).
In many aspects qualitative
properties of different types of motion are similar in four and five
dimensions. Both four- and five-dimensional metrics are
algebraically special and allow two families of principle null
congruences. In both cases these congruences are geodesic. The
principal null rays are also eigen-vectors of the tensor
$\xi_{\mu;\nu}$. However, there are some differences. While in
four dimensions the principle null congruences are shear free in five
dimensions this is not the case.
Also, in four dimensions  there exist stable circular orbits around the
rotating black hole, while for  five dimensional
Myers-Perry black hole they are absent, at least in the `equatorial' planes.

It would be interesting to generalize these results to
rotating black holes in arbitrary number of dimensions.

\bigskip

\vspace{12pt} {\bf Acknowledgments}:\ \  This work was partly
supported  by  the Natural Sciences and Engineering Research
Council of Canada. The authors are grateful to the Killam Trust
for its financial support.

\bigskip

\appendix
\section{Optical scalars for null congruences in a higher dimensional
space-time}
\label{ap}
\setcounter{equation}0

Consider a congruence of null geodesics in $n$-dimensional
space-time
and let $l^{\mu}$ be a tangent vector, then
\be\n{a.1}
l^{\mu}l_{\mu}=0\, ,\hspace{1cm}
l^{\mu}_{\ ;\nu}\, l^{\nu}=0\, ,\hspace{1cm}
l^{\mu}_{\ ;\nu}\, l_{\mu}=0\, .
\ee
Denote by $\Sigma$ \ \ $(n-1)$-dimensional null plane which is spanned
be vectors $z^{\mu}$ obeying the condition $z^{\mu}\, l_{\mu}=0$. Let us
choose another null vector $n^{\nu}$, normalized by the condition
$n^{\mu}\, l_{\mu}=-1$, and denote by $S$ \ $(n-2)$-dimensional subspace
of $\Sigma$ formed by vectors which are orthogonal to both null vectors.
Denote by $h_{\mu\nu}$ a projector onto $S$
\be\n{a.2}
h_{\mu\nu}=g_{\mu\nu}+2\, l_{(\mu}\, n_{\nu)}\, .
\ee
One has
\be \n{a.3}
h_{\mu\nu}l^{\nu}=h_{\mu\nu}n^{\nu}=0\, ,\hspace{1cm}
h_{\mu}^{\ \ \alpha}h_{\alpha\nu}=h_{\mu\nu}\, .
\ee

As usual let us decompose $l_{\mu;\nu}$ as follows
\be\n{a.4}
l_{\mu;\nu}=l_{[\mu;\nu]}+\sigma_{\mu\nu}+{1\over n-2}\vartheta\,
h_{\mu\nu}\, ,
\ee
where
\be\n{a.5}
\vartheta=l^{\alpha}_{\ \ ;\alpha}=h^{\alpha\beta}\, l_{\alpha;\beta}\,
,\hspace{1cm}
\sigma_{\mu\nu}=l_{(\mu;\nu)}-{1\over n-2}\vartheta \, \, h_{\mu\nu}\, .
\ee
The parameters of twist, $\hat{\rho}$, and shear, $\sigma$, are defined as
follows
\be\n{a.6}
\hat{\rho}^2=l_{[\mu;\nu]}l^{[\mu;\nu]}\, ,\hspace{1cm}
\sigma^2=\sigma_{\mu\nu}\sigma^{\mu\nu}=h^{\alpha\mu}\, h^{\beta\nu}\,
l_{(\alpha;\beta)} l_{(\mu;\nu)}-{1\over n-2}\vartheta^2\, .
\ee
It should be emphasized that the choice of the null vector $n^{\mu}$
contains ambiguity $n^{\mu}\to n^{\mu}+z^{\mu}$ where $z^{\mu}$ is any
vector from $\Sigma$. Under this transformation $h_{\mu\nu}\to
h_{\mu\nu}+2l_{(\mu}\, z_{\nu)}$, while the expansion $\vartheta$, twist
$\hat{\rho}$ and shear $\sigma$ remain invariant.

In a space-time with two principle null geodesic congruences
$L_{\pm}^{\mu}$ the natural choice of the projector $h_{\mu\nu}$
is
\be\n{a.7}
h_{\mu\nu}=g_{\mu\nu}-2\alpha {L_{+}}_{(\mu}\, {L_{-}}_{\nu)}\,
,\hspace{1cm}
\alpha=({L_{+}}_{\mu}\, {L_{-}}^{\mu})^{-1}\, .
\ee

For the 5-dimensional metric (\ref{Kerr5D}) and the principle null
vectors (\ref{PNG}) one has
\be\n{a.8}
\alpha= {-\Delta \over 2x \rho^2}\, ,
\ee
\be\n{a.9}
h^{\mu\nu}={1\over x\rho^2}\times
\left[ \begin{array}{ccccc}
    x(a^2\sin^2\theta+b^2\cos^2\theta)+a^2b^2 & 0 & 0 & -a(x+b^2) &
    -b(x+a^2) \\
    0 & 0 & 0 & 0 & 0 \\
    0 & 0 & x & 0 & 0 \\
    -a(x+b^2) & 0 & 0 & {\displaystyle x+b^2\sin^2\theta \over
    \displaystyle \sin^2\theta} & ab  \\
    -b(x+a^2) & 0 & 0 & ab & {\displaystyle x+a^2\cos^2\theta\over
    \displaystyle \cos^2\theta}
\end{array}\right]\, .
\ee
(The matrix $ h^{\mu\nu}$ is written in the basis
$(t,x,\theta,\phi, \psi)$.)



\end{document}